\documentclass[reprint,prx,aps,amsmath,amssymb,floatfix,superscriptaddress,longbibliography]{revtex4-2}

\usepackage{mathtools,amsmath}
\usepackage{graphicx} 
\usepackage[breaklinks=true,colorlinks,citecolor=teal,linkcolor=teal,urlcolor=teal]{hyperref}
\usepackage{physics}
\usepackage{siunitx}
\usepackage{bbold}
\usepackage{soul}
\usepackage{bm}
\usepackage[normalem]{ulem}
\usepackage{times}
\usepackage{enumerate}  
\usepackage{float}
\usepackage{amssymb}
\usepackage{xcolor} 
\usepackage{subfigure}

\begin{document}

\title{Generation of entanglement and magic via continuous homodyne monitoring of a qubit pair}

\author{Debmalya Das}
\affiliation{Department of Physics, School of Advanced Sciences, VIT-AP University,
Beside AP Secretariat, Amaravati 522241, Andhra Pradesh, India}

\author{Giuseppe Magnifico}
\affiliation{Dipartimento Interuniversitario di Fisica, Universit\`a di Bari, I-70126 Bari, Italy}
\affiliation{INFN, Sezione di Bari, I-70125, Bari, Italy}

\author{Maria Maffei}
\affiliation{Universit\'e de Lorraine, CNRS, LPCT, F-54000 Nancy, France}


\begin{abstract}
Continuous weak measurements of quantum systems are of great relevance in quantum foundations and applications. They can be achieved by probing the quantum system of interest by repeatedly measuring an auxiliary system weakly coupled to it. Here we study two qubits coupled in different points to a common one-dimensional electromagnetic field and simultaneously monitored in the right- and left-propagating output channels by homodyne detection. Using a collision-model description, we derive an analytical Stochastic Master Equation (SME) governing the resulting diffusive quantum trajectories, including the interference between the two measurement channels. For nonlinear functions of the quantum state, such as entropies, averages over quantum trajectories generally differ from the corresponding quantities evaluated on the unconditional state. Through this mechanism, we show that continuous monitoring generates entanglement, absent in the unconditional dynamics, and enhances quantum magic in the qubit pair during the decay. Both resources can be tuned through the optical phase accumulated between the qubits and the phases of homodyne local oscillators. 
Our results establish continuous homodyne monitoring of multiple emitters as a tunable mechanism for generating quantum resources.
\end{abstract}
\maketitle

\section{Introduction}\label{secI}
Continuously monitored quantum systems~\cite{Gisin1984, Busch1984, Gisin1993, Wiseman2010, Jacobs2014, Jordan_Siddiqi_2024} have long been used as playgrounds for exploring fundamental aspects of quantum theory, such as quantum contextuality~\cite{dressel2010}, steering~\cite{Wiseman_steering2012}, and weak values~\cite{ Aharonov1988,  Duck1989, Wiseman_weak_values2002, Huard2014, maffei2022_wigner, Stevens2022}. They also find diverse applications in many relevant branches of quantum technology, such as feedback control~\cite{Vijay2012, Patti2017, Albarelli2024, Karmakar2026}, quantum entanglement generation and protection~\cite{Williams2008, Lewalle2017, Lewalle2020, Lewalle2021}, quantum state-preparation and magic state distillation~\cite{Karmakar2026}, characterization and certification of time-dependent gate Hamiltonians~\cite{Siva2023}, quantum  state estimation~\cite{Das2014, Das2015, Das2017}, quantum metrology~\cite{Rossi2020, Duan2025}, and dynamical phase transitions~\cite{Turkeshi2021}.

Continuous monitoring of a microscopic quantum system can be achieved with repeated measurements of a second extended system, or reservoir, to which it is weakly coupled. A typical experimental platform where this can be observed is CircuitQED, where superconducting qubits or qudits are weakly coupled to the microwave electromagnetic fields propagating in the circuits~\cite{Murch2013, Huard2014,Weber2016,Stevens2022, Steinmetz2022}. In this framework, the unmonitored dynamics of the microscopic system typically obeys a Master Equation (ME)~\cite{Brun2002, Gambetta2008}. The reservoir is repeatedly measured, giving rise to a string of output values, or \textit{record}. Different records  correspond to different evolutions of the system, i.e. different \textit{quantum trajectories} (QTs). The latter are solutions of a Stochastic Master Equation (SME). Taking the ensemble average of the system's states over many trajectories is equivalent to taking a trace over the reservoir and hence gives the solution of the ME~\cite{Wiseman2010}. 

\begin{figure}[t!] 
\includegraphics[width=1.00\linewidth]{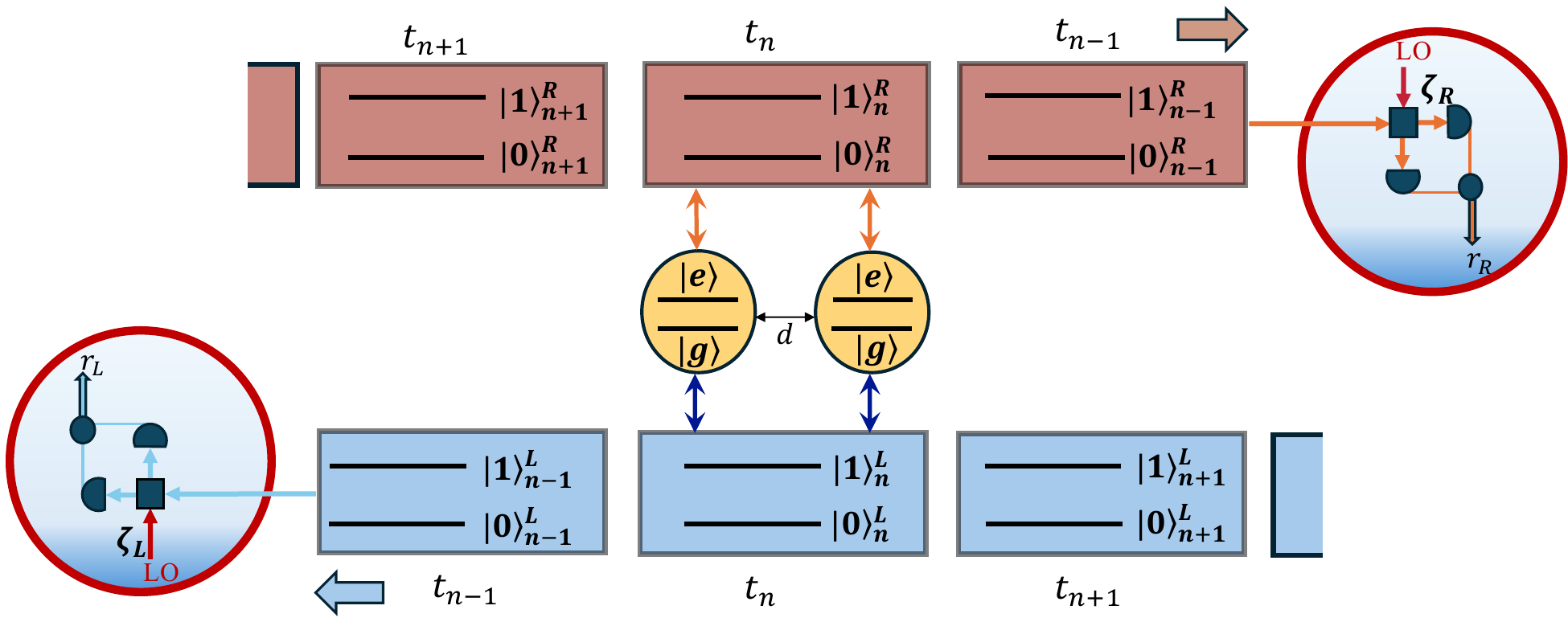}
\caption{\textbf{Collision model of continuous homodyne monitoring of a qubit pair.} Right- and left-propagating electromagnetic fields are modeled as two reservoirs of probe-units which interact with the qubits shortly one after the other. The units are subsequently measured with balanced homodyne detectors placed at the emitters' sides and the readouts, $r_L$ and $r_R$, are recorded. The local oscillators (LO) used in right and left homodyne detectors have phases $\zeta_R$ and $\zeta_L$ with respect to the field emitted by the qubits. The probe units are labeled according on their arrival time at the qubits position and interact with the two qubits at the same time. This is a valid description when the time-of-flight of the field between the emitters, $t_d$, is much shorter than their excited states' lifetimes, $\gamma^{-1}$, such that the duration of each collision, $\Delta t$, controlling the ``size'' of the units, can be taken greater than $t_d$ but still far smaller than $\gamma^{-1}$. Due to the the weakness of the qubits-field coupling each unit is treated as an effective two-level system.} 
    \label{fig:1}
\end{figure}

Here we consider a pair of identical qubits weakly coupled with an electromagnetic field confined to propagate in one-dimension (1D). The qubits are coupled to the field at two different points separated by a distance $d$. We consider that the field's time-of-flight between the two points, $t_d$, is much smaller than the qubits' excited state lifetime $\gamma^{-1}$; within this hypothesis the reduced dynamics of the qubit pair obeys a Lindblad Master Equation~\cite{Cilluffo2020}. On the other hand, the frequency $\omega_0$ of each qubit is much greater than $\gamma$ and hence, in general, the optical phase $\varphi=t_d\omega_0$ takes finite values which determine the properties of the decay dynamics, yielding super and sub radiance~\cite{Gross1982,Facchinetti2016,Facchi2016}, or directional emission~\cite{maffei2024directional}. As the 1D electromagnetic field can propagate in both directions, the qubits are coupled to two reservoirs, made of right- and left-propagating photons, respectively, which can be weakly measured simultaneously and independently. In particular, we analyze the case where the right and left reservoirs are simultaneously monitored via homodyne by using two local oscillators with different phases, $\zeta_R$ and $\zeta_L$. We derive the SME which specifies the monitored dynamics of the two qubit state starting from the Hamiltonian of the joint qubit pair-field system. Our derivation exploits the fact that the joint qubit-field unitary dynamics can be mapped onto a collision model~\cite{Ciccarello_2017, Ciccarello20221} and generalizes the one derived for a single emitter coupled to a single reservoir presented in Ref.~\cite{Gross2018}. In the collision model, right and left electromagnetic reservoirs are regarded as sequences of probe-units that interact sequentially and briefly with the qubits and are subsequently measured one by one (see the schematics in Fig.~\ref{fig:1}). As expected, the SME of double homodyne detection depends on the set of parameters $(\varphi, \zeta_R,\zeta_L)$.

For quantities depending non-linearly on the qubits' state, as von Neumann and Rényi entropy, statistical averages over quantum trajectories differ from values computed on the unmonitored state~\cite{Ravindranath2023, O'Dea2024, Li2025, Delmonte2025, Pinol2026}. Then, averaging over many quantum trajectories, we show that two relevant quantum resources, entanglement and magic~\cite{Leone22}, are generated and enhanced along the monitored dynamics of the qubit pair. We display how these resources depend on the values of the tunable parameters $(\varphi, \zeta_R,\zeta_L)$. The present results can be generalized to other measurements schemes and to settings featuring more than two emitters increasing the number of tunable control parameters and hence obtaining richer phase-diagrams of the quantum resources generated by the monitoring.

The paper is structured as follows. In Sec.~\ref{secII} we describe the collision model of the qubits dynamics and use it to derive the Master Equation ruling the unmonitored evolution. In Sec.~\ref{seciii} we add the homodyne measurements to the considered setting and use the collision model to derive the Stochastic Master Equation describing the monitored dynamics of the qubit pair, this original derivation is the central technical result of the paper. In Sec.~\ref{seciv} we present our main results displaying how quantum entanglement and magic arise in the two qubit state due to the continuous monitoring and how they can be tuned changing the phases of homodyne detectors and the optical phase acquired by the electromagnetic field propagating between the qubits. Finally in Sec.~\ref{secv} we draw the conclusions of the work.

\section{System and Model}\label{secII}

We consider a pair of identical qubits coupled to the same one-dimensional waveguide at different points, $x_1$ and $x_2$, at a distance $d$ from each other. The bare Hamiltonian of qubit $j\in\{1,2\}$ is $H_j= \omega_0 \sigma_{+}^{(j)}\sigma_{-}^{(j)}$, where $\sigma_{-}^{(j)}=|g_j\rangle\langle e_j|$, and $\sigma_{+}^{(j)}=\sigma_{-}^{(j)\dag}$ with $e$ and $g$ labeling the excited and the ground state, respectively. For notation shortness, we will denote the states of the tensor product emitter basis as $\ket{ee}$, $\ket{eg}$, $\ket{ge}$, and $\ket{gg}$; also, without loss of generality, we will set $x_1=-d/2$ and $x_2=d/2$.

The electromagnetic field propagates along the waveguide with linear dispersion relation in the relevant bandwidth around $\omega_0$, with constant group velocity $v_g$. We can then separate photons propagating towards right and towards left by introducing the bosonic operators $a_{R}(\omega)$ and $a_{L}(\omega)$ which annihilate, respectively, one right-propagating (R) photon of frequency $\omega=\omega_0 + v_g k $, and one left-propagating (L) photon of frequency $\omega=\omega_0 - v_g k $, with $k$ being the photon wave-number taken positive for the right-propagating, and negative for the left-propagating ones. Hence the field's bare Hamiltonian reads $H_{f}=\int d\omega~\omega \left(a_L^\dagger(\omega) a_L (\omega)+ a_R^\dagger (\omega) a_R (\omega)\right)$, and the light-matter coupling reads
\begin{align}\label{V_Schroedinger}
V&=\sqrt{\frac{\gamma}{4\pi}}\sum_{j=1,2} \int d\omega~\sigma_{-}^{(j)} \left[e^{i \omega_0 x_j/v_g} a^{\dagger}_{R}(\omega)\right.\\ \nonumber
&\left. + e^{-i \omega_0 x_j/v_g} a^{\dagger}_{L}(\omega)\right] +\mathrm{H.c.}.
\end{align}
Let us notice that the above interaction Hamiltonian assumes the  rotating-wave and flat-coupling approximations and hence it provides an accurate description of the light-matter interaction only in the regime where the coupling is weak and the bare qubits frequency $\omega_0$ is much greater than $\gamma$  ~\cite{Gardiner1985Input}.

Moving in the interaction picture with respect to the bare Hamiltonians of qubits and field, $H_{0}=H_{1}+H_{2}+H_{f}$, the light-matter coupling Hamiltonian of Eq.~\eqref{V_Schroedinger} becomes 
\begin{align}\label{eq:V_int}
    V_{I}(t)=& \sqrt{\frac{\gamma}{2}}\Biggl\lbrace \sigma_{-}^{(1)} \left[ e^{i \varphi/2} b^{\dagger}_{R}\left(t\right)+   e^{-i \varphi/2} b^{\dagger}_{L}\left(t\right)\right]   \\ \nonumber 
    & +\sigma_{-}^{(2)} \left[e^{-i \varphi/2} b^{\dagger}_{R}\left(t- t_d \right)+ e^{i \varphi/2} b^{\dagger}_{L}\left(t+t_d\right)\right] \Biggr\rbrace + \mathrm{H.c.}
\end{align}
Here $t_d=d/v_g$ is the time of flight of the light between the qubits, $\varphi=\omega_0 t_d$ is the corresponding optical phase, and
$b_{\ell}(t)$, with $\ell\in\lbrace R,L\rbrace$, are the so-called quantum noise operators~\cite{gardinerzoller}, defined as $b_{\ell}(t)=(2\pi)^{-1/2}\int d\omega e^{-i(\omega-\omega_0) t}a_{\ell}(\omega)$ and verifying bosonic commutation relations $[b_{\ell}(t),b^{\dagger}_{m}(t')]=\delta_{\ell,m}\delta(t-t')$.
\\
\\
The delay between the quantum noise operators, $t_d$, in the general case, makes the emitters dynamics no longer solvable with a Lindblad ME rendering necessary more complex computational approaches such as tensor networks simulations~\cite{Magnifico2025,Capurso2025}. Nevertheless, when $t_d\gamma \ll1$, one can take the approximation $b_{\ell}(t\pm t_d)\approx b_{\ell}(t)$ remaining in the regime of validity of the Lindblad ME, which is the case considered in this paper. Let us stress that the condition $\omega_0\gg\gamma$ implies that $\varphi$ takes in general finite values.

\subsection{Collision model}

Following refs.~\cite{Cilluffo2020} and~\cite{maffei2024directional}, we treat the emitters-field dynamics with a quantum collision model. We define a discrete time variable $t_n= n\Delta t$ taking the interval $\Delta t$ such that $t_d<\Delta t\ll \gamma^{-1}$. Then we define discrete annihilation operators acting on field's time-bin units: 
\begin{equation}
	b_{\ell,n} = \frac{1}{\sqrt{\Delta t}} \int_{n \Delta t}^{(n+1)\Delta t} dt~b_{\ell}(t),
\end{equation}
which satisfy $[b_{\ell,n},b^{\dagger}_{m,n'}]=\delta_{\ell,m}\delta_{n,n'}$. The condition $\Delta t>t_d$ implies that, at time $t_n$, both qubits are coupled (collide) with the $n$-th time-bin unit of the field (see Fig.~\ref{fig:1}). The unitary evolution operator evolving the joint light-matter system during the interval $[n\Delta t,(n+1)\Delta t]$, up to the first order in $\gamma\Delta t$ is expressed through the Magnus expansion:
\begin{align}\label{Magnus}
    U_n &= \mathbb{1} - i \int_{n \Delta t}^{(n+1) \Delta t} \dd t \ V_I(t) \\ \nonumber
    &- \frac{1}{2} \int_{n \Delta t}^{(n+1) \Delta t} \dd t \int_{n \Delta t}^{t} \dd t' \ [V_I(t), V_I(t')] +\mathcal{O}(\gamma\Delta t).
\end{align}
The first term of the expansion reads:
\begin{equation}\label{eq:V_discreteTime}
 \int_{n \Delta t}^{(n+1)\Delta t}dt V_{I}(t) =\sqrt{\gamma \Delta t}\sum_{\ell=R,L} \left( \mathcal{J}^{\,}_{\ell} b^{\dagger}_{\ell,n} + h.c. \right)\equiv V_{n}\Delta t,
\end{equation} 
where we defined the jump operators associated to right and left emission:
\begin{align} 
    \mathcal{J}_{R}  &= \sqrt{\frac{1}{2}} \left(e^{i\varphi/2}\sigma_{-}^{(1)} + e^{-i\varphi/2}\sigma_{-}^{(2)} \right),
    \end{align}
    \begin{align}
    \mathcal{J}_{L} & = \sqrt{\frac{1}{2}} \left(e^{-i\varphi/2} \sigma_{-}^{(1)} + e^{i\varphi/2} \sigma_{-}^{(2)} \right) . \label{eq:vertexJ2_sm}
\end{align}
The second term of~\eqref{Magnus} instead gives an effective exchange interaction between the qubits
\begin{align}
     &-\frac{i}{2\Delta t} \int_{n \Delta t}^{(n+1) \Delta t} \dd t \int_{n \Delta t}^{t} \dd t' \ [V_I(t), V_I(t')] \\ \nonumber
    &= \frac{\gamma}{2} \sin (\varphi) (\sigma_{+}^{(1)} \sigma_{-}^{(2)} + \sigma_{-}^{(1)} \sigma_{+}^{(2)})\equiv H_{e},
\end{align}
yielding a simpler expression of $U_n$:
\begin{align}\label{eq:unitary_2qubits}
    U_n= \mathbb{1}- i \Delta t \left(V_{n} + H_e\right)-\frac{\Delta t^2}{2} V_{n}^2 +\mathcal{O}(\gamma\Delta t).
\end{align}
As we are interested in the spontaneous decaying dynamics of the two emitters, in the rest of the paper, we will consider that all time-bins modes prior to the collision are in vacuum states. In this case the Lindblad ME can be easily retrieved from the trace over the $n$-th field's unit:
\begin{align}\label{eq:ME}
\rho(t_{n+1})&=\text{Tr}_{n}\left\lbrace U_{n} \ket{0_{n}}\bra{0_{n}} \rho(t_n)U^{\dagger}_n \right\rbrace \rightarrow\\ \nonumber
 \Delta \rho &=-i \left[H_{e},\rho\right]\Delta t+\gamma \Delta t \left(\mathcal{D}_{\mathcal{J}_{R}}[\rho]+
\mathcal{D}_{\mathcal{J}_{L}}[\rho]\right),
\end{align}
where $\Delta\rho \equiv \rho(t_{n+1})-\rho(t_n)$ is the change of the reduced qubits state between time $t_n$ and $t_{n+1}$, and $\mathcal{D}_{X}[\rho]= X\rho X^\dagger-\frac{1}{2}\left(\rho X^\dagger X + X^\dagger X \rho\right)$ is the Lindblad superoperator.



\section{Homodyne measurements and quantum diffusive trajectories}\label{seciii}

Cutting $U_n$ at first order in $\gamma \Delta t$ implies that, after the interaction, each unit can contain either 0 or 1 excitation, and hence it enacts as an effective two-level system (see Fig.~\ref{fig:1}). This feature greatly simplifies the expressions of Kraus operators of homodyne measurement that we derive in this section, following Ref.~\cite{Gross2018}. In the collisional picture, the continuous homodyne detection corresponds to projecting the $n$-th right- (left-) propagating unit, after its interaction with the emitters, onto the eigenstates of the qadrature $\mathcal{X}^{n}_R$ ($\mathcal{X}^{n}_L$) which read   
\begin{eqnarray}
   \ket{\pm_R}_{n}&=&\frac{1}{\sqrt{2}}\left(\ket{0_R}_n \pm e^{i\zeta_R}\ket{1_R}_n\right),
   \end{eqnarray}
   \begin{eqnarray}
    \ket{\pm_L}_{n}&=&\frac{1}{\sqrt{2}}\left(\ket{0_L}_n \pm e^{i\zeta_L}\ket{1_L}_n\right).
\end{eqnarray}
The four possible measurement outcomes $\ket{\pm_{R}}_n\otimes \ket{\pm_{L}}_n$ define a set of Kraus operators $\{K_{lm}\}$, with $l,m=\pm$, which act on the emitters:
\begin{equation}
    K_{lm}=\bra{l_L,m_R}U_n\ket{0_{R},0_{L}}_{n}
\end{equation}
From eq.~\eqref{eq:unitary_2qubits}, it follows that

 \begin{eqnarray}\label{eq:Krauss}
&&K_{lm} = \frac{1}{2}\left\lbrace \mathbb{1} -\frac{\Delta \tau}{2}\left(\mathcal{J}_L^\dagger \mathcal{J}_L+\mathcal{J}_R^\dagger \mathcal{J}_R\right)-iH_{e}\Delta \tau \right.\\ \nonumber
 &&+\Delta\tau \left[(\epsilon_{l-}+\epsilon_{l+})(\epsilon_{m+}+\epsilon_{m-})X_LX_R\right] \\\nonumber
 &&+\left. \sqrt{\Delta\tau}\left[(\epsilon_{l-}+\epsilon_{l+})X_L+(\epsilon_{m+}+\epsilon_{m-})X_R\right]\right\rbrace,
\end{eqnarray}
with $\Delta \tau \equiv \gamma \Delta t$, $\epsilon_{+-}=1$, $\epsilon_{-+}=-1$, and $\epsilon_{++}=\epsilon_{--}=0$ being the Levi-Civita symbols, and $X_{L}\equiv e^{-i\zeta_L}\mathcal{J}_{L}$ and $X_{R} \equiv e^{-i\zeta_R}\mathcal{J}_{R}$ being defined as two phase-shifted jump operators. 

The associated POVMs $E_{lm}=K^\dagger_{lm} K_{lm}$, to the first order in $\Delta \tau$, are given by
\begin{eqnarray}\label{eq:POVM_h2em}
    E_{lm} = \frac{1}{4}&&\left \lbrace \mathbb{1} +\Delta L (X_L+X_L^\dagger)+\Delta R (X_R +X_R^\dagger) \right.\\ \nonumber
    &&\left.+\Delta R~\Delta L [(X_L+ X_L^\dagger)X_{R}+ X_R^\dagger (X_L+ X_L^\dagger)]\right\rbrace,
\end{eqnarray}
where we defined the stochastic variables corresponding to the measurements outputs, $\Delta L=(\epsilon_{l-}+\epsilon_{l+})\sqrt{\Delta \tau}$, and $\Delta R=(\epsilon_{m-}+\epsilon_{m+})\sqrt{\Delta \tau}$. 

From the above operators, we can derive the variation of the reduced qubits state through a single step of the monitored dynamics conditioned on the instantaneous outcome $(l,m)$: 
\begin{equation}
    \Delta \rho\vert_{(l,m)}= \frac{K_{lm}\rho K^\dagger_{lm}}{\text{Tr}\lbrace \rho E_{lm}\rbrace}-\rho.
\end{equation}
Notice that the denominator is the probability of obtaining the output $(l,m)$ at a certain time $t_n$, and satisfies the normalization relations: $\sum_{l,m=\{\pm\}} \text{Tr}\lbrace \rho E_{lm} \rbrace= 1$, $\text{Tr} \lbrace \rho E_{++}\rbrace + \text{Tr} \lbrace \rho E_{--}\rbrace=1/2=\text{Tr} \lbrace \rho E_{+-}\rbrace+ \text{Tr} \lbrace \rho E_{-+}\rbrace$.

Plugging Eq.~\eqref{eq:Krauss} and~\eqref{eq:POVM_h2em} in the above expression, and retaining terms till the first order in $\Delta \tau$ we get
\begin{widetext}
\begin{eqnarray}
    \Delta \rho\vert_{(l,m)} &=& \left(\Delta L-\Delta \tau \text{Tr}\left\lbrace \rho(X_L+X_L^\dagger)\right\rbrace \right)\mathcal{H}_{X_L}[\rho]+\left(\Delta R-\Delta \tau \text{Tr}\left\lbrace \rho(X_R+X_R^\dagger)\right\rbrace \right)\mathcal{H}_{X_R}[\rho]+\Delta L~ \Delta R~\mathcal{G}_{X_L,X_R}[\rho] \nonumber\\
    &+&\Delta t\left[-i \left[H_{e},\rho\right]+\gamma \left(\mathcal{D}_{\mathcal{J}_L}[\rho]+\mathcal{D}_{\mathcal{J}_R}[\rho]\right)\right],
\label{eq:change_lm}
\end{eqnarray}
with
\begin{eqnarray}
\mathcal{H}_{X}[\rho]=X\rho + \rho X^\dagger - \rho \text{Tr}\left[\rho\left(X+X^\dagger\right)\right],
\end{eqnarray}
    \begin{eqnarray}
\mathcal{G}_{X_L,X_R}[\rho]&&= X_{L}(X_{R}\rho +\rho X_{R}^{\dagger}) + (X_{R}\rho +\rho X_{R}^{\dagger})X_{L}^{\dagger} -\rho \text{Tr}\lbrace (X_{L}+X_{L}^{\dagger})(X_{R}\rho +\rho X_{R}^{\dagger})\rbrace\\ \nonumber
&&-\text{Tr}\lbrace (X_{L}+X^{\dagger}_L) \rho \rbrace \mathcal{H}_{X_R}[\rho] -  \text{Tr}\lbrace (X_{R}+X^{\dagger}_R) \rho \rbrace \mathcal{H}_{X_L}[\rho].
    \label{eq:defn_G}
\end{eqnarray}
\end{widetext}
The last term of Eq.~\eqref{eq:change_lm} corresponds to the unconditional Master Equation [Eq.~\eqref{eq:ME}] and, as will become clear in a moment, it is the only term surviving upon averaging over many measurements runs. 
We can compare Eq.~\eqref{eq:change_lm} with the homodyne SME for a single emitter and a single monitored decay channel (let's say R) as reported in Ref.~\cite{Gross2018} that, with our notation, reads $\Delta \rho\vert_{r} = \left(\Delta R -\Delta \tau \text{Tr}\left\lbrace \rho(X_R+X_R^\dagger)\right\rbrace \right)\mathcal{H}_{X_R}[\rho]+\Delta t\mathcal{D}_{\mathcal{J}_R}[\rho]$. It is clear that the term $\mathcal{G}_{X_{L},X_{R}}[\rho]$ features an interference between right and left measurements as a consequence of the simultaneous detections. 
\begin{figure}[t]
    \centering
    \includegraphics[width=0.99\linewidth]{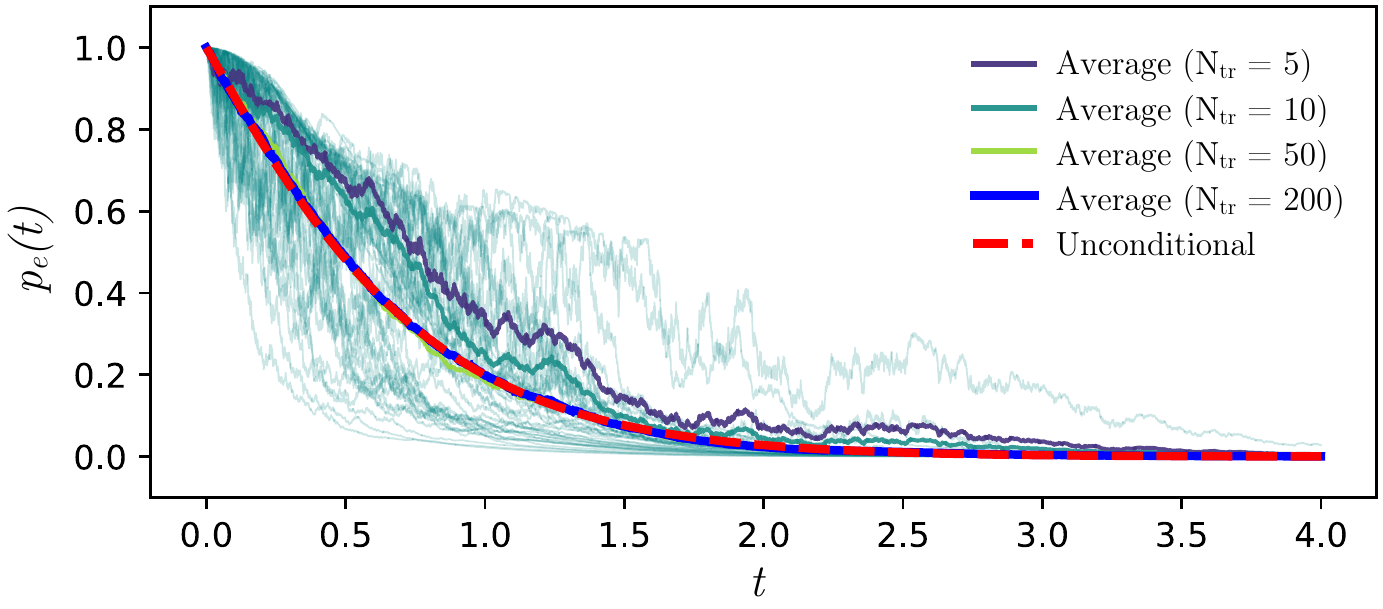}
    \caption{\textbf{Probability of excitation of either one of the qubits, along different trajectories, starting from the doubly excited state.} Excitation probability along various trajectories obtained from Eq.~\eqref{eq:SME} with $\varphi=2\pi$ and $\zeta_{R}=0=\zeta_{L}$. The average over 200 trajectories (blue continuous line) is identical to the unmonitored probability (red dashed) as given by the Lindblad ME.}
    \label{fig:qubit_dynamics}
\end{figure}

We compute the first moments of the stochastic variables $\Delta L$ and $\Delta R$:
\begin{eqnarray}
    \mathbb{E}[\Delta L] &=& \sqrt{\Delta \tau}\sum_{m=\lbrace\pm\rbrace}\text{Tr}\lbrace\rho (E_{+m} -E_{-m})\rbrace\\ \nonumber
    &=&\Delta \tau \text{Tr}\left\lbrace \rho\left(X_L + X^\dagger_L\right)\right\rbrace,\\
  \mathbb{E}[\Delta R] &=& \sqrt{\Delta \tau}\sum_{l=\lbrace\pm\rbrace}\text{Tr}\lbrace\rho (E_{l+} -E_{l-})\rbrace\\ \nonumber
    &=&\Delta \tau \text{Tr}\left\lbrace \rho\left(X_R + X^\dagger_R\right)\right\rbrace.
\end{eqnarray}
And from the first moments we define the so called \textit{homodyne innovations}~\cite{wiseman2009quantum}:
\begin{eqnarray}
    \Delta \mathcal{I}_{\mathcal{H}L}=\Delta L - \mathbb{E}[\Delta L],\\
    \Delta \mathcal{I}_{\mathcal{H}R}=\Delta R - \mathbb{E}[\Delta R].
\end{eqnarray}
The homodyne innovations have zero mean and have variance equal to $\Delta \tau$, using these quantities we can reformulate Eq.~\eqref{eq:change_lm} as
\begin{widetext}
    \begin{equation}
   \Delta \rho(t) = \Delta \mathcal{I}_{\mathcal{H}L}\mathcal{H}_{X_L}[\rho]+\Delta \mathcal{I}_{\mathcal{H}R}\mathcal{H}_{X_R}[\rho]+\Delta \mathcal{I}_{\mathcal{H}L}\Delta \mathcal{I}_{\mathcal{H}R}\mathcal{G}_{X_L,X_R}[\rho] + \Delta t\left[-i \left[H_{e},\rho\right]+\gamma \left(\mathcal{D}_{\mathcal{J}_L}[\rho]+\mathcal{D}_{\mathcal{J}_R}[\rho]\right)\right]. 
    \label{eq:SME}
\end{equation}
\end{widetext}
The above form is the discrete-time version of the stochastic master equation ruling the dynamics of the two qubits under continuous homodyne detections of their right- and left-propagating emitted fields. In the continuous time limit, the Innovation $\Delta \mathcal{I}_{\mathcal{H}R}$ are replaced by $\sqrt{\gamma}dW_L$ and $\sqrt{\gamma}dW_R$, respectively, which are mutually independent Wiener processes, satisfying $\mathbb{E}[dW_{L/R}]=0$ and $\mathbb{E}[dW^2_{L/R}]=dt$, by the rules of Ito calculus~\cite{Jacobs2006, Wiseman2010, Gross2018, Jordan_Siddiqi_2024}. In this limit, the qubits quantum trajectories can be generated assuming that the values of $dW_{R(L)}$ follow a Gaussian distribution of zero mean and variance $dt$. 

Fig.~\ref{fig:qubit_dynamics} shows the excitation probability of any of the two qubits along different trajectories generated by the SME [Eq.~\eqref{eq:SME}] starting from the initial state $\ket{ee}$. The plot shows clearly how individual trajectories may diverge substantially from the unmonitored evolution, yet averaging over a growing number of trajectories, $p_{e}$ progressively converges to its unconditional value (red dashed decreasing exponential) which is reached with a near perfect fidelity even when the average is taken over a sample of a few hundreds of trajectories. Notice that this is true for any quantity which depends linearly on the qubits state, while it does not hold for non-linear functions of $\rho$ as we'll see in the next section.

\begin{figure*}[t]
     \includegraphics[width=0.78\linewidth]{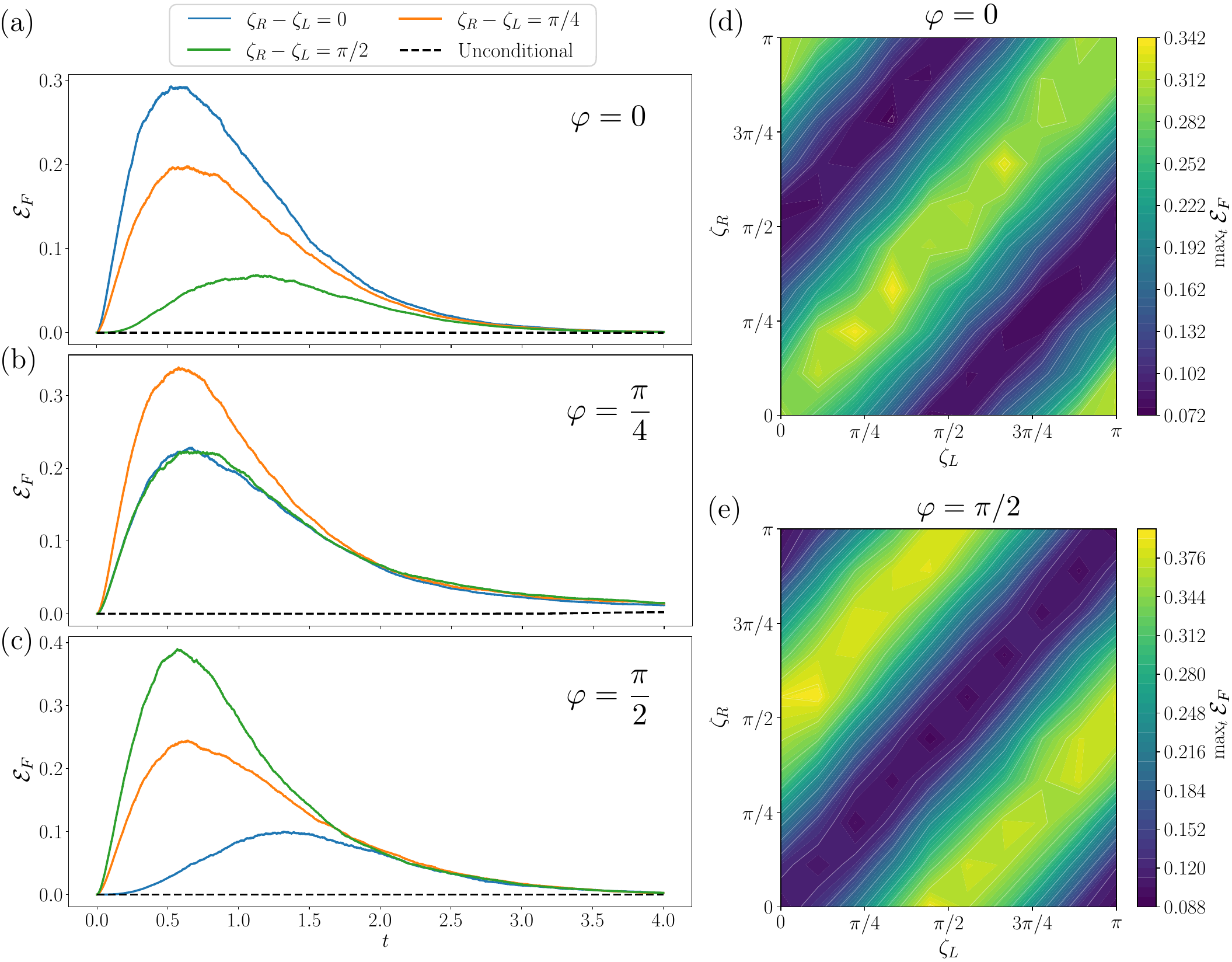}
    \caption{\textbf{Entanglement of formation of the state of the monitored qubit pair averaged over thousand trajectories.} a-c Entanglement of formation as a function of time for fixed values of the parameters $\varphi$ and $\zeta_R-\zeta_L$ starting from the state $\ket{ee}$. In the unmonitored dynamics the state has zero entanglement (dotted line), in the monitored dynamics averaged over thousand trajectories the Entanglement of formation is equal to the von Neumann entropy of either one of the two emitters as their joint state is pure. d-e Maximum value of the $\mathcal{E}_{F}$ reached along the monitored decay for fixed $\varphi$ varying $\zeta_R$ and $\zeta_L$. The plots show that the qubits state is more entangled when $\Phi=(\zeta_R- \zeta_L)-\varphi=n \pi$ with $n\in \mathbb{N}_0$, and has minimum entanglement when $\Phi=(2n+1) \pi/2$.   }
    \label{fig:panel_eof}
\end{figure*}

\section{Results}\label{seciv}

The qubits and the probe-units form a closed system. In this framework, along the unmonitored dynamics dictated by the Lindblad ME [Eq.~\eqref{eq:ME}], the purity loss of the two qubit state comes solely from the entanglement that the latter build with the reservoir. Measuring each probe immediately after its interaction with the emitters erases the entanglement with the reservoir in each step. This reflects on the fact that, unlike the ME [Eq.~\eqref{eq:ME}], the SME [Eq.~\eqref{eq:SME}] preserves the purity of the joint qubits state. In this section, we quantify two resources that are present in the pure state of the monitored qubits: entanglement and quantum magic. Importantly, in pure states, both resources can be quantified by the use of entropies, respectively, von Neumann and Rényi entropy~\cite{Wootters1998, Leone22}. Since the latter are non-linear functions of the two qubit state, their average values over many quantum trajectories do not converge to the values that they would take in the unmonitored dynamics. We show that the monitored evolution produces, on average, two qubit states having entanglement and quantum magic until their decay is completed. We display how these quantities depend on the tunable parameters $(\varphi,\zeta_R,\zeta_L)$. 

\subsection{Entanglement}

Entanglement of formation ($\mathcal{E}_F$) can be used to quantify the entanglement between two qubits in an arbitrary mixed state~\cite{Wootters1998, PhysRevLett.128.040501}. Computing the $\mathcal{E}_{F}$ of the unmonitored state of the qubits given by Eq.~\eqref{eq:ME}, one finds zero entanglement at each time of the decay starting from $\ket{ee}$, for any value of $\varphi$. When the state is pure, $\mathcal{E}_{F}$ coincides with the von Neumann entropy of either one of the two parties, $S[\rho_1]=S[\rho_2]$, with $S[\rho]=-\text{Tr}\lbrace \rho \text{log}[\rho]\rbrace$, and hence its maximum allowed value is $\text{log}(2)$.  We computed the $\mathcal{E}_{F}$ (or equivalently the von Neumann entropy) of the pure state of the qubit pair along many quantum trajectories and then we averaged the obtained values. We find non-zero entanglement during all the decay dynamics until the ground state is reached. Figure~\ref{fig:panel_eof} shows the $\mathcal{E}_{F}$ averaged over thousand trajectories generated by the SME in Eq.~\eqref{eq:SME} varying the parameters $(\varphi,\zeta_{R},\zeta_{L})$. Importantly it displays that the entanglement depends on the angle $\Phi=(\zeta_R- \zeta_L)-\varphi$: the state is more entangled when $\Phi=n \pi$ with $n\in \mathbb{N}_0$, and has minimum entanglement when $\Phi=(2n+1) \pi/2$. We can connect the build-up of entanglement in the monitored qubit pair with the extraction of which-path information from the measurements outputs: the more which-path information can be extracted from the data, the less entangled the qubits wavefunction is after measurement. This trade-off between entanglement and which-path information is expected in general in any interferometric setting in quantum mechanics~\cite{Wineland93,EnglertPRL}. In the present case, $\Phi$ is the relative phase between the homodyne signals of right and left detectors. When $\Phi$ is an integer multiple of $\pi$, the probability distributions of right and left measurements overlap. When $\Phi=(2n+1) \pi/2$, right and left distributions are maximally distinguishable and so the data can be used to tell if the emission happened towards right or towards left (the data carry maximal which-path information). In agreement with the general theory, the former case gives the highest $\mathcal{E}_{F}$, the latter gives the lowest $\mathcal{E}_{F}$, while values of $\Phi$ in between, corresponding to an intermediate amount of which-path information, give an intermediate amount of entanglement. The described mechanism was already reported in Ref.~\cite{Lewalle2021} within a simpler model where the qubits were not coupled to the same propagating field, but to individual output channels mixed afterwards in a beam splitter. The present analysis then includes the additional element of the optical phase $\varphi$. Furthermore it can be generalized to a number $N$ of emitters coupled to the propagating field obtaining $N-1$ optical phases $\varphi_i$ to tune the generation of resources.

\subsection{Quantum Magic}

\begin{figure*}[t!]
\includegraphics[width=0.78\linewidth]{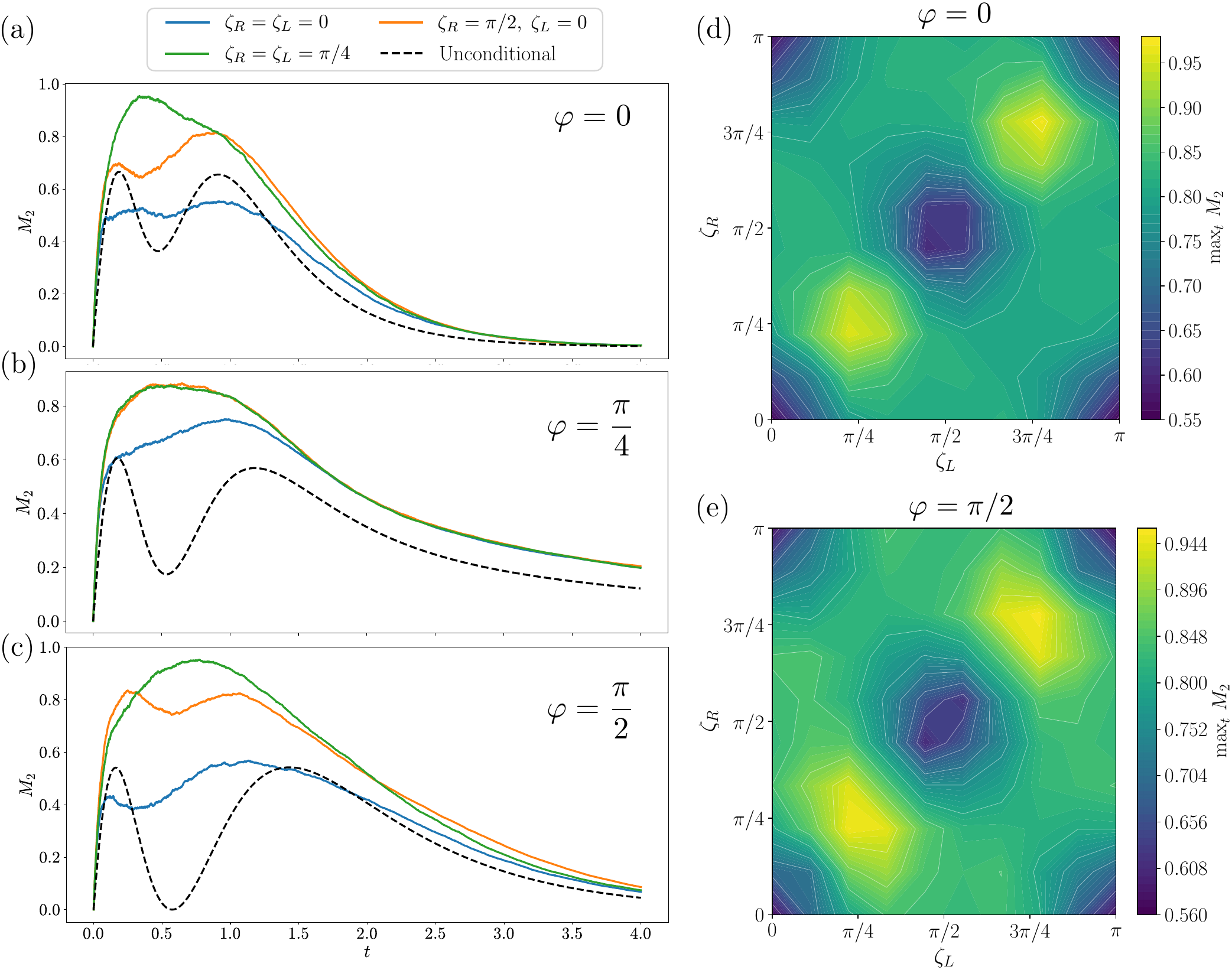}
\caption{\textbf{Stabilizer 2-Rényi entropy of the state of the monitored qubit pair averaged over thousand trajectories.} a-c Stabilizer 2-Rényi entropy as a function of time for fixed values of the parameters $\varphi$ and $\zeta_R, \zeta_L$ starting from the state $\ket{ee}$. For all parameters values the magic increases already after the first step of the monitored evolution showing that the Kraus operators of Eq.~\eqref{eq:Krauss} are non-Clifford gates.  d-e Maximum value of the $M_2$ reached along the monitored decay for fixed $\varphi$ varying $\zeta_R$ and $\zeta_L$: the magic is maximal when $\zeta_R= \zeta_L=\pi/4 + n \pi$ with $n\in \mathbb{N}_0$, and minimal when $\zeta_R=\zeta_L=n\pi$.}
    \label{fig:panel_magic}
\end{figure*}

Quantum magic, or non-stabilizerness, quantifies the extent to which a quantum state lies outside the set of stabilizer states, i.e. those preparable from computational basis states using only Clifford operations \cite{Leone22}. The Gottesman-Knill theorem tells us that Clifford circuits acting on stabilizer states can be efficiently simulated on a classical computer, regardless of the amount of entanglement they generate \cite{gottesman1998heisenbergrepresentationquantumcomputers, PhysRevA.70.052328, 2011350.2011356}. Therefore, the preparation and distillation of magic states \cite{Bravyi2005, PhysRevLett.102.110502, PhysRevLett.118.090501} is needed for quantum computational advantage. Very recently, in Ref.~\cite{Karmakar2026}, continuous monitoring along with feedback has been employed to achieve magic state distillation.
Magic has also emerged as a useful way of characterizing the non-classical complexity of quantum states \cite{Leone22, Veitch_2014, Hou_2026} and dynamics, in particular monitored quantum dynamics \cite{Russomanno_2025, 31sq-k4m3, PRXQuantum.5.030332, PhysRevResearch.6.L042030, Turkeshi_2025, tirrito2025magicphasetransitionsmonitored}, the class of processes considered here. 
Specializing to two qubit systems, stabilizer states are those reachable by applying Clifford gates to the state $\ket{ee}$ ~\cite{Gottesman1998}. A representative set of Clifford gates comprises the Hadamard gate, the Phase gate, and the CNOT gate. A two-qubit state carries magic when it cannot be written as a single stabilizer state, but only as a linear combination of several of them, the minimal number of which defines its magic rank~\cite{Knill2004, Bravyi2005}. Notably, Bell states are stabilizer states and therefore have zero quantum magic. 

Ref.~\cite{Leone22} demonstrated that the magic of a pure state of two qubits is quantified by its stabilizer 2-Rényi entropy:
\begin{align}\label{eq:magic_def}
M_{2}(\rho)=-\text{log}_{2}\left[\left( \sum_{i,j=0,x,y,z} \frac{\text{Tr}\left\lbrace  \sigma_i\otimes\sigma_j \rho \right\rbrace^2 }{4}\right)^2\right]-2,
\end{align}
with $\lbrace \sigma_{0},\sigma_{x},\sigma_{y},\sigma_{z}\rbrace$ being the Pauli matrices. $M_2$ takes maximum value equal to $2$. Along each trajectory the monitored state is pure, so $M_{2}$ directly quantifies its magic. The unmonitored state is generally mixed, therefore, to compare the two cases, we need the mixed-state generalization of Eq.~\eqref{eq:magic_def}. The latter, introduced in the same Ref.~\cite{Leone22}, reads $\tilde{M_2}(\rho) = M_2(\rho) - S_2(\rho)$, where $S_2(\rho) = -\textrm{log}_2\textrm{Tr}(\rho^2)$ is the 2-Rényi entropy. As $\tilde{M_2}$ reduces to Eq.~\eqref{eq:magic_def} for pure states, the magic of monitored and unmonitored states are quantified with the same measure. Figure ~\ref{fig:panel_magic} shows the stabilizer 2-Rényi entropy of the monitored qubit-pair averaged over thousand trajectories starting from the state $\ket{ee}$, together with the corresponding unconditional values. Both monitored and unmonitored states have some magic since the very beginning of the evolution for any values of $(\varphi,\zeta_R,\zeta_{L})$, yet their behaviors as a function of time are rather different. The magic of the unconditional state (dashed line) exhibits pronounced oscillations whose minima decrease when increasing the optical phase $\varphi$. The trajectory-averaged magic of the monitored state oscillates much less and remains substantially larger along all the decay dynamics until the relaxation is completed. Its dependence on the optical phase is then weaker than the unmonitored case, while it shows a clear pattern as a function of the homodyne phases. Panels d-e display the maximum of the trajectory-averaged $M_2$ over the decays versus $(\zeta_R,\zeta_{L})$ at fixed $\varphi$: the magic is maximal when $\zeta_R= \zeta_L=\pi/4 + n \pi$ with $n\in \mathbb{N}_0$, and minimal when $\zeta_R=\zeta_L=n\pi$.
Hence continuous monitoring enhances the average magic of the state and makes it tunable.

\section{Conclusions and outlooks}\label{secv}

We showed that weak measurements of a qubit pair via continuous monitoring of their emitted field can generate and enhance valuable quantum resources in their joint state. In the proposed setting, implementable within state-of-the-art superconducting circuits, the amount of such resources can be controlled by three tunable parameters: the optical phase acquired by the field in the propagation between the qubits, and the two phases of the homodyne detectors detecting simultaneously the right- and the left-propagating output fields. Using a collision model, we derived the analytical expression of the stochastic master equation ruling the qubits trajectories during such continuous monitoring. We then used a large sample of trajectories to quantify the resources, i.e. entanglement and quantum magic, present in the monitored state and to display how they can be tuned with the three control parameters. The present study can be straightforwardly generalized to chains of more than two qubits or qudits coupled weakly to electromagnetic fields, and to different kinds of field's detection schemes such for instance single or number-resolved photon-counting, and heterodyne. In particular, the extension of the proposed scheme to a chain of $N$ emitters would generate states whose resources are controlled by $(N-1)+2$ independent parameters: the $N-1$ inter-qubits optical phases plus the two phases introduced by the homodyne detectors. Interfacing such a rich playground with numerical optimization methods has the potential to produce a wide variety of multi-qubits states useful for computation or communication.  

\section{Acknowledgments}
The authors thank Mario Collura, Cyril Elouard, Cosmo Lupo and Marco Di Liberto for their valuable inputs. G.M. acknowledges support from INFN through the projects  ``QUANTUM'' and ``NPQCD'', from the Italian funding within the  ``Budget MUR - Dipartimenti di Eccellenza 2023-2027'' - Quantum Sensing and Modelling for One-Health (QuaSiModO), and from Projects PN-RIC Q-SUD (Q-SUD B99H26000380007) and PN-RIC QUANTAS (SYNERGIA B99H26000410007). We acknowledge computational resources provided by the University of Bari and the INFN cluster ReCaS~\cite{ReCaS}.

\bibliography{biblio}

\end{document}